\documentclass[twocolumn,letter]{jpsj3}

\usepackage{graphicx}
\usepackage{dcolumn}
\usepackage{bm}
\usepackage{color}
\usepackage{txfonts}
\def\vector#1{{\boldsymbol{#1}}}

\def\vd{{\vector d}}

\def\vk{{\vector k}}

\def\vx{{\vector x}}
\def\vy{{\vector y}}
\def\vz{{\vector z}}

\def\kB{{k_{\rm B}}}

\def\SrRuO{\mbox{${\rm Sr_2RuO_4}$}}

\def\eq.#1{Eq.~(\ref{#1})}
\def\eqs.#1{Eqs.~(\ref{#1})}

\def\refeq#1{(\ref{#1})}

\def\Hc2{{H_{\rm c2}}}
\def\difHc2{{H'_{\rm c2}}}

\def\*red*#1{{\color{red}{#1}}}

\newcommand\Equation[2]{
\begin{equation}\label{#1} 
#2
\end{equation}
}

\newcommand\Equationnoeqn[2]{
\begin{equation*}\label{#1} 
#2
\end{equation*}
}

\title{
Vestigial Van Hove Singularity and Higher-Temperature Superconducting \\ 
Phase Induced by Perpendicular Uniaxial Pressures \\ 
in Quasi-Two-Dimensional Superconductors 
}

\author{Hiroshi Shimahara}

\inst{
Graduate School of Advanced Science and Engineering, Hiroshima University, 
Higashi-Hiroshima 739-8530, Japan 
}

\recdate{July 6, 2020}

\abst{
We examine quasi-two-dimensional superconductors near half-filling 
under uniaxial pressures perpendicular to conductive layers 
(hereafter called perpendicular pressures). 
It is a natural conjecture that 
the perpendicular pressure decreases 
the transition temperature $T_{\rm c}$ 
because it increases the interlayer electron hopping energy $t_z$, 
which weakens the logarithmic enhancement in the density of states 
due to the two-dimensional Van Hove singularity. 
It is shown that, contrary to this conjecture, 
the perpendicular pressure can significantly enhance $T_{\rm c}$ 
in systems off half-filling 
before it decreases $T_{\rm c}$, 
and the strength of the enhancement 
significantly depends on the pairing symmetry. 
When the indices 
d, ${\rm d}'$, cz, and sz 
are defined 
for the basis functions 
$\gamma_{\rm d} \propto \cos k_x - \cos k_y$, 
$\gamma_{\rm d'} \propto \sin k_x \sin k_y$, 
$\gamma_{\rm cz} \propto \cos k_z$, 
and $\gamma_{\rm sz} \propto \sin k_z$, 
respectively, 
it is shown that 
for s-, d-, cz-, and cz-d-wave pairing, 
$T_{\rm c}$ steeply increases with increasing $t_z$ 
near a cusp at a certain value of $t_z$. 
On the other hand, 
for p-, cz-p-, sz-p-, and ${\rm d}'$-wave pairing, 
$T_{\rm c}$ is almost unaffected by $t_z$. 
For sz- and sz-d-wave pairing, 
$T_{\rm c}$ exhibits a broad and weak peak. 
Here, for example, 
the cz-d-wave state is an interlayer spin-singlet d-wave state 
with an order parameter proportional 
to $\gamma_{\rm cz} \gamma_{\rm d}$. 
The enhancement in $T_{\rm c}$ is the largest for this state 
and the second largest for the d-wave pairing 
and interlayer spin-singlet (cz-wave) pairing. 
These results may explain recent observations in 
${\rm Sr_2RuO_4}$ under perpendicular pressures. 
A comparison between the theoretical and experimental results 
indicates that the p-, cz-p-, and sz-p-wave states, 
including chiral states, and the ${\rm d}'$-wave state 
are the most likely candidates for the intrinsic 1.5-K phase, 
and 
the d-, cz-d-, and cz-wave states 
are the most likely candidates for 
the 3-K phase induced by the perpendicular pressure. 
The cz-p- and sz-p-wave states are 
interlayer spin-triplet and interlayer spin-singlet p-wave states 
with horizontal line nodes, respectively. 
}

\begin{document}
\sloppy
\maketitle


{\it Introduction} --- 
It has been frequently pointed out that 
the logarithmic Van Hove singularity in the density of states 
can enhance the superconducting transition temperature $T_{\rm c}$ 
in quasi-two-dimensional systems 
near half-filling~\cite{Hir86,Mar97,Shi87,Shi88,Mar91}. 
Despite the long history of research on this mechanism, 
the effect of interlayer electron motion 
on this mechanism has not been frequently examined, 
probably because 
the interlayer hopping energy $t_z$ increases 
the dimensions of the system 
and removes the logarithmic singularity~\cite{Mar91}; 
hence, $t_z$ seems to decrease $T_{\rm c}$. 
In this study, however, we illustrate that 
$t_z$ can significantly enhance the density of states and $T_{\rm c}$.

The effect of $t_z$ on superconductivity might have been 
observed recently in a real material. 
The compound {\SrRuO} is 
a quasi-two-dimensional superconductor 
with tetragonal symmetry~\cite{Mae94,Mac03}, 
and its pairing symmetry has been 
under debate since the discovery of 
superconductivity in this compound~\cite{Mae94,Mac03,Mae12}. 
To clarify the pairing symmetry, 
the application of uniaxial pressures in various directions 
can be a useful tool~\cite{Hic14,Kit10,Tan12,Tan15,Ima20}. 
\mbox{Hicks et al.} examined 
shifts of $T_{\rm c}$ 
due to symmetry-breaking in-plane strains~\cite{Hic14} 
and found that a $[1,0,0]$ strain significantly 
enhances $T_{\rm c}$. 
They argued that the orthorhombic distortion enhances 
the density of states owing to the Van Hove singularity~\cite{Shi87}, 
and this can be the origin of the enhancement in 
$T_{\rm c}$~\cite{Hic14}. 
Although this phenomenon under an in-plane pressure 
is not directly related to the effect of $t_z$ 
that we examine below, 
it suggests that 
the Van Hove singularity can enhance $T_{\rm c}$ by an appreciable amount 
in this compound. 
Before their study, 
\mbox{Kittaka et al.} examined {\SrRuO} 
under uniaxial pressures in the $[0,0,1]$ direction, 
which is perpendicular to conductive layers 
(hereafter called perpendicular pressures)~\cite{Kit10}. 
Their study suggests that a perpendicular pressure induces superconductivity 
with an onset $T_{\rm c}$ above 3\,K, 
while interestingly, 
the transition to the intrinsic superconducting phase remains 
with its transition temperature $T_{\rm c} \approx 1.5~{\rm K}$ 
appearing to be unaffected. 
In contrast to the in-plane strains, 
the distortion in this direction does not change 
the symmetry of the system 
and weakens the enhancement in the density of states 
due to the Van Hove singularity. 
Nevertheless, the observed phenomena can be explained 
by a scenario based on a vestigial Van Hove singularity, 
which will be explained below.

The tetragonal quasi-two-dimensional system can be modeled by 
the one-particle electron energy 
\Equation{eq:Q2Depsilon}
{
     \epsilon_{\vk} 
     = \epsilon_{\vk_{\parallel}}^{\parallel} 
       - 2 t_{z} \cos k_z , 
     }
where 
${
     \epsilon_{\vk_{\parallel}}^{\parallel} 
     = - 2 t (\cos k_x + \cos k_y) 
     }$ 
with $\vk_{\parallel} = (k_x,k_y)$ 
and the lattice constants $a$, $b$, and $c$ have been absorbed into 
the definitions of the momentum components 
$k_x$, $k_y$, and $k_z$, respectively. 
When $t_z = 0$, the saddle points of 
$\epsilon_{\vk_{\parallel}}^{\parallel}$ 
at $\vk_{\parallel} = (\pm \pi,0)$ and $(0,\pm \pi)$ 
give rise to the Van Hove singularity. 
A perpendicular pressure increases $t_z$ 
and removes the singularity. 
We denote the electron density per site and the chemical potential 
as $n$ and $\mu$, respectively. 
When we apply the theory to {\SrRuO}, 
the dispersion in \eq.{eq:Q2Depsilon} is a simplified model; 
however, 
the model near half-filling can simulate 
the physical situation of the $\gamma$ band in this compound, 
in which the Fermi surface is near the saddle points~\cite{Hic14}. 
We use units in which $\hbar = \kB = t  = 1$.


{\it Density of states} --- 
The mechanism by which $t_z$ enhances the density of states 
can be interpreted as follows. 
The density of states defined by 
\Equationnoeqn{eq:rhodef}
{
     \rho(\epsilon) 
     = \frac{1}{N} \sum_{\vk} \delta (\epsilon - \epsilon_{\vk}) 
     }
is expressed as 
\Equation{eq:rhodefinrhopara}
{
     \rho(\epsilon) 
     = \int_{-\pi}^{\pi} \frac{d k_z}{2 \pi} 
       \rho_{\parallel}(\epsilon + 2 t_z \cos k_z) 
     }
with 
\Equationnoeqn{eq:rho2D}
{
     \rho_{\parallel} (\epsilon) 
     \equiv \int \frac{d^2 k}{(2 \pi)^2} 
         \delta (\epsilon - \epsilon_{\vk}^{\parallel}) 
     }
being the density of states of 
the square lattice system~\cite{Noterhopara}. 
Here, $N$ denotes the number of sites. 
$\rho_{\parallel}(\epsilon)$ 
diverges logarithmically at $\epsilon = 0$, 
and 
$\rho_{\parallel} \approx (2\pi^2 t)^{-1}\ln (16t/|\epsilon|)$ 
for $|\epsilon| \ll t$. 
When $t_z \ne 0$, 
the integration over $k_z$ in \eq.{eq:rhodefinrhopara} 
removes this divergence, 
and when $t_z$ increases, 
the peak height $\rho(0)$ decreases, as expected; 
however, 
because $\int \rho (\epsilon) d \epsilon = 1$ is a constant, 
the suppression of the peak height leads to 
an increase in $\rho(\epsilon)$ 
in some other regions of $\epsilon$. 
It is evident from \eq.{eq:rhodefinrhopara} that 
when $|\mu| \leq 2 t_z$, 
the contribution to $\rho(\mu)$ 
from the electron states near $k_z = \pm \arccos(- \mu/2t_z)$ is large 
because of the logarithmic enhancement in $\rho_{\parallel}$. 
It is verified that $\partial \rho/\partial \epsilon = 0$ 
for $|\epsilon| \leq 2 t_z$, 
which implies that the top of the vestigial peak is a plateau.

\begin{figure}[htbp]
\begin{center}
\begin{tabular}{c}
\\ 
\includegraphics[width=5.5cm]{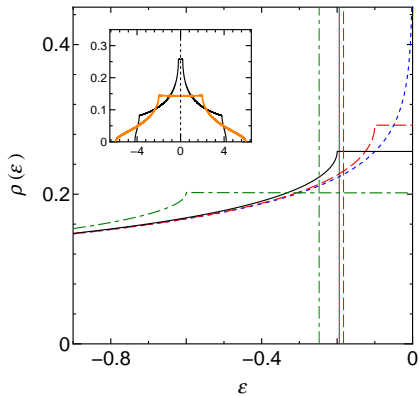}
\\[-12pt]
\end{tabular}
\end{center}
\caption{
(Color online) 
Densities of states and the Fermi levels. 
The curves and thin vertical lines show 
$\rho(\epsilon)$ and $\epsilon = \mu$, 
respectively, the intersections of which yield $\rho(\mu)$ 
for various values of $t_z$. 
The red dashed, black solid, 
and green dot-dashed curves and lines show 
the results for $t_z = 0.05$, $0.1$, and $0.3$, respectively. 
The blue short-dashed curve shows $\rho(\epsilon)$ 
for $t_z = 0$. 
The inset shows overall profiles of $\rho(\epsilon)$. 
The orange thick solid curve shows 
$\rho(\epsilon)$ for $t_z = 1$. 
}
\label{fig:rho}
\end{figure}

Figure~\ref{fig:rho} 
illustrates how the perpendicular pressure 
enhances the density of states $\rho(\mu)$ at the Fermi level 
when the system is nearly half-filled. 
The curves show $\rho(\epsilon)$, 
and the thin vertical lines indicate 
$\epsilon = \mu$ for $n = 0.9$. 
The logarithmic singularity in $\rho(\epsilon)$ disappears 
for any finite $t_z$, 
and a plateau appears~\cite{Mar91}. 
The density of states at the Fermi level $\rho(\mu)$ 
increases as $t_z$ increases from 0 
when the system is not half-filled. 
For example, 
$\rho(\mu) \approx 0.231$ for $t_z = 0.05$, 
whereas 
$\rho(\mu) \approx 0.257$ for $t_z=0.1$, 
as shown by the red dashed and black solid curves, 
respectively. 
When $t_z$ increases further, $\rho(\mu)$ decreases. 
For example, $\rho(\mu) \approx 0.202$ for $t_z = 0.3$. 
In s-wave superconductors, 
the increase and decrease in $\rho(\mu)$ immediately result in 
an increase and a decrease in $T_{\rm c}$, respectively.


{\it Superconductivity} --- 
In anisotropic superconductors, 
$T_{\rm c}$ is a function of an effective density of states, 
in which the momentum dependence of the order parameter 
is incorporated. 
The pairing interaction is expanded as 
\Equation{eq:Vkk}
{
     V_{\vk\vk'} = - \sum_{\alpha} {\bar g}_{\alpha} 
       \gamma_{\alpha}(\vk)
       \gamma_{\alpha}(\vk') , 
     }
where 
$\alpha$ is the index of the basis function 
and ${\bar g}_{\alpha}$ is the coupling constant for 
the $\alpha$-wave state. 
The functions $\gamma_{\alpha}(\vk)$ 
are orthonormal bases~\cite{Shi87}, 
which satisfy 
\Equationnoeqn{eq:normalization}
{
     \frac{1}{N} \sum_{\vk} \gamma_{\alpha}(\vk) 
                            \gamma_{\alpha'}(\vk) 
     = \delta_{\alpha \alpha'} . 
     }
The pressure affects the values of ${\bar g}_{\alpha}$; 
however, 
we leave the effect of the change in ${\bar g}_{\alpha}$ 
for future research 
and focus on the effect of the change in the density of states.

The order parameter is expanded as 
\Equationnoeqn{eq:Deltaexp}
{
     \Delta_{\vk} 
     = \sum_{\alpha} \Delta_{\alpha} \gamma_{\alpha} (\vk) , 
     }
and the linearized gap equations are 
\Equationnoeqn{eq:Tceq}
{
     \Delta_{\alpha} 
       = \frac{{\bar g}_{\alpha} }{N} \sum_{\vk} \sum_{\alpha'} 
         \gamma_{\alpha}(\vk) 
         W(\xi_{\vk}) 
         \gamma_{\alpha'}(\vk) \Delta_{\alpha'} , 
     }
where
$W(\xi_{\vk}) = {\tanh (\beta \xi_{\vk}/2)}/{2 \xi_{\vk}}$. 
Because of the symmetry of the system, 
these equations are decoupled into subsets by the pairing symmetries. 
When the pairing state is not a mixed-symmetry state, 
the order parameter $\Delta_{\vk}$ is a linear combination of 
basis functions with the same symmetry, 
which is expressed as 
\Equation{eq:Deltaexpsym}
{
     \Delta_{\vk} 
     = \sum_{\alpha \in S_{\lambda}} 
         \Delta_{\alpha} \gamma_{\alpha} (\vk) , 
     }
where $S_{\lambda}$ is a set of $\alpha$ values 
such that all $\gamma_{\alpha}$ have the same symmetry $\lambda$. 
As a consequence of the superposition, 
the order parameter of the most stable state is localized near 
the Fermi surface in momentum space~\cite{Notespps}, 
reflecting the range of interaction of the order of 
$v_{\rm F}/\omega_{\rm c}$, 
which is much larger than the lattice constants, 
where $v_{\rm F}$ denotes the Fermi velocity. 
In this paper, 
we simplify the problem by retaining 
a single principal basis function $\gamma_{\alpha}$ 
for each pairing symmetry 
and restrict the range of interaction by introducing 
the cutoff energy $\omega_{\rm c}$ 
instead of superposing many basis functions to localize 
$\Delta_{\vk}$ near the Fermi surface. 
Hence, we retain a single $\alpha$ in the summations 
in \eqs.{eq:Vkk} and \refeq{eq:Deltaexpsym} 
and replace $\gamma_{\alpha}(\vk)$ with 
$     C \, \theta (\omega_{\rm c} - |\xi_{\vk}|) 
       \, \gamma_{\alpha}( \vk ) $, 
where $C$ is a normalization constant 
and $\xi_{\vk} \equiv \epsilon_{\vk} - \mu$. 
The equation for $T_{\rm c}$ is 
\Equationnoeqn{eq:Tceqsimple}
{
     1 = \frac{g_{\alpha}}{N} \sum_{\vk} 
         W(\xi_{\vk}) \theta(\omega_{\rm c} - |\xi_{\vk}|) 
           [\gamma_{\alpha}(\vk)]^2 , 
     }
and when $\omega_{\rm c} \ll t$, 
we obtain 
\Equation{eq:Tcexpress}
{
     T_{\rm c} = \frac{2 e^{\gamma}}{\pi} 
       \omega_{\rm c} e^{- 1/\lambda_{\alpha}} , 
     }
where 
$\lambda_{\alpha} = g_{\alpha} \rho_{\alpha}(\mu)$, 
$g_{\alpha} \equiv {\bar g}_{\alpha} C^2$, 
and 
$\gamma = 0.57721 \cdots$ is the Euler's constant. 
Here, 
$\rho_{\alpha}(\epsilon)$ is 
the effective density of states for $\alpha$-wave pairing, 
which is expressed as 
\Equationnoeqn{eq:rhoeffective}
{
     \rho_{\alpha}(\epsilon) \equiv 
         \int \frac{d^3 k}{(2 \pi)^3} 
         \delta(\epsilon - \epsilon_{\vk}) 
         [{\gamma}_{\alpha}(\vk)]^2 . 
     }

We adopt 
$     {\gamma}_{{\rm p}_x}(\vk) 
     = \sqrt{2} \sin k_x$ 
and 
    $\gamma_{{\rm p}_y}(\vk) 
     = \sqrt{2} \sin k_y$
as the principal basis functions of 
the ${\rm p}_x$- and ${\rm p}_y$-wave states, respectively. 
These states are degenerate in the tetragonal system, 
and they and any superposition of them, 
for example, 
the chiral ${\rm p}_x \pm i {\rm p}_y$ wave states, 
have the same transition temperature. 
Hence, as far as $T_{\rm c}$ is concerned, 
we simply call them the p-wave states. 
Among them, the one with the lowest free energy occurs below $T_{\rm c}$, 
and presumably, the chiral states have the lowest free energy 
because they are full-gap states. 
We adopt 
$\gamma_{\rm s} = 1$ 
and 
$\gamma_{{\rm d}} = \cos k_x - \cos k_y$, 
respectively, 
as the principal bases of the s- and d-wave states. 
For the ${\rm d}_{xy}$-wave state, 
we adopt 
$\gamma_{{\rm d}_{xy}} = 2 \sin k_x \sin k_y$. 
We also examine interlayer pairing 
between adjacent layers~\cite{Has00}, 
for which the order parameter has a factor $\cos k_z$ or $\sin k_z$. 
Hence, the resultant order parameter is a product of 
$\cos k_z$ or $\sin k_z$ 
and an in-plane basis, such as 
$\gamma_{\rm s}$, 
$\gamma_{\rm p}$, and $\gamma_{\rm d}$. 
For example, when the latter is the s-wave function, 
i.e., ${\hat \gamma_{\rm s}} = 1$, 
the principal bases are 
\Equationnoeqn{eq:interlayergamma}
{
     \gamma_{\rm cz}(\vk) = \sqrt{2} \cos k_z , ~~~~  
     \gamma_{\rm sz}(\vk) = \sqrt{2} \sin k_z , 
     }
where we defined the indices cz and sz 
for $\cos k_z$ and $\sin k_z$, respectively. 
The cz-wave state is a spin singlet, 
whereas the sz-wave state is a spin triplet. 
For the d- or p-wave in-plane states, 
the principal bases are 
\Equationnoeqn{eq:interlayer_d_and_p}
{
     \begin{split}
     \gamma_{\rm cz-d}
       & = \gamma_{\rm cz}
           \gamma_{\rm d} , ~~~~~ 
     \gamma_{\rm sz-d}
         = \gamma_{\rm sz}
           \gamma_{\rm d} , \\ 
     \gamma_{\rm cz-p}
       & = \gamma_{\rm cz} 
           \gamma_{\rm p} , ~~~~~ 
     \gamma_{\rm sz-p}
         = \gamma_{\rm sz} 
           \gamma_{\rm p} . \\ 
     \end{split}
     } 
The cz-d- and sz-p-wave states are spin singlets, 
whereas the sz-d- and cz-p-wave states are spin triplets.

\begin{figure}[htbp]
\begin{center}
\begin{tabular}{c}
\\ 
\includegraphics[width=5.5cm]{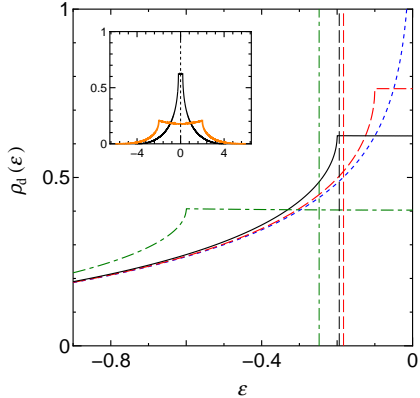}
\\[-12pt]
\end{tabular}
\end{center}
\caption{
(Color online) 
Effective density of states $\rho_{\rm d}(\epsilon)$ 
for the d-wave pairing with $\Delta_{\vk} \propto \cos k_x - \cos k_y$. 
The legend of this figure is the same as that of Fig.\,\ref{fig:rho}. 
}
\label{fig:rhod}
\end{figure}

Figure~\ref{fig:rhod} shows that 
the effective density of states $\rho_{\rm d}(\mu)$ 
at the Fermi level is enhanced by the same mechanism as 
that for $\rho(\mu)$, 
and the enhancement in $\rho_{\rm d}(\mu)$ 
is much larger than 
that in 
$\rho_{\rm s}(\mu) = \rho(\mu)$ 
because 
$[\gamma_{\rm d}(\vk)]^2$ is large near 
the saddle points at $(k_x,k_y) = (\pm \pi,0)$ and $(0,\pm \pi)$. 
This example illustrates that 
the enhancement effect of the present mechanism 
significantly depends on the pairing symmetry. 
For a comparison between different pairing symmetries, 
we evaluate $T_{\rm c}$ under the condition that 
the values of $T_{\rm c}$ at $t_z = 0$ are equated. 
For explicit evaluations, we adopt specific values 
$n = 0.9$ and $\omega_{\rm c} = 300~{\rm K}$ 
and assume that $T_{\rm c} \approx 1.5~{\rm K}$ 
at $t_z = 0$~\cite{Notegalpha}.

\begin{figure}[htbp]
\begin{center}
\begin{tabular}{c}
\\ 
\includegraphics[width=6.5cm]{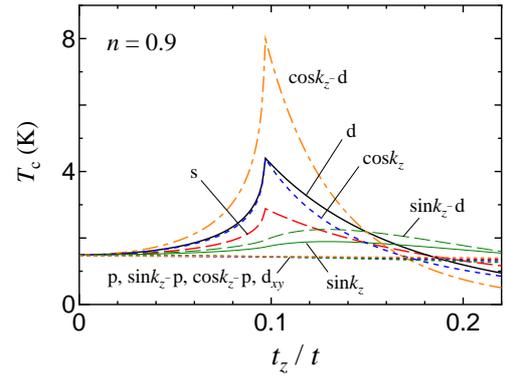}
\\[-12pt]
\end{tabular}
\end{center}
\caption{
(Color online) 
Transition temperatures for various pairing symmetries 
when $T_{\rm c} \approx 1.5\,{\rm K}$ 
at $t_z = 0$~\cite{Notegalpha}. 
The black solid 
and red dashed 
curves present the results for 
the d- and s-wave states, respectively. 
The 
red, 
blue, 
orange, 
and 
green thick dotted 
curves present the results for 
the p-, cz-p-, sz-p-, 
and ${\rm d}_{xy}$-wave states, respectively. 
Most parts of the dotted curves overlap. 
The blue short-dashed 
and orange dot-dashed 
curves present the results for 
the cz- and cz-d-wave states, respectively. 
The green thin solid 
and dashed 
curves present the results for 
the sz-s- and sz-d-wave states, respectively. 
}
\label{fig:Tc_tz0Tc15}
\end{figure}

\begin{figure}[htbp]
\begin{center}
\begin{tabular}{c}
\includegraphics[width=6.5cm]{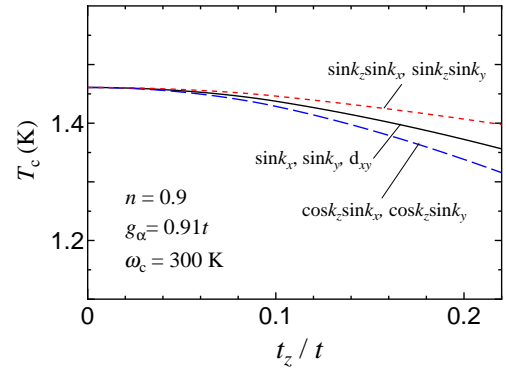}
\\[-12pt]
\end{tabular}
\end{center}
\caption{
(Color online) 
Transition temperatures when 
$n = 0.9$, 
$\omega_{\rm c} = 300~{\rm K}$, 
and $g_{\alpha} = 0.91 t$ 
for $\alpha = {\rm p}$, cz-p, sz-p, 
and ${\rm d}_{xy}$. 
}
\label{fig:inplane_pw}
\end{figure}

The results are shown in Fig.\,\ref{fig:Tc_tz0Tc15}, 
and it is found that the enhancement in $T_{\rm c}$ is 
the largest and the next largest 
for the cz-d-wave state and the d- and cz-wave states, respectively. 
For these three states and s-wave states, 
$T_{\rm c}$ increases steeply near a cusp at a certain value of $t_z$. 
For the interlayer sz- and sz-d-wave states, 
$T_{\rm c}$ exhibits a broad peak. 
For p-, sz-p-, cz-p-, and ${\rm d}_{xy}$-wave states, 
$T_{\rm c}$ changes little when $t_z$ increases. 
(Strictly speaking, $T_{\rm c}$ decreases slightly 
as shown in Fig.\,\ref{fig:inplane_pw}.) 
This originates from the fact that 
the order parameters of these states vanish 
at the saddle points of 
$\epsilon_{\vk_{\parallel}}^{\parallel}$ 
because of the in-plane bases proportional to 
$\sin k_x$, $\sin k_y$, or $\sin k_x + i \sin k_y$. 
Note that this result holds for any p-wave states 
because every term of the order parameters of the p-wave states 
is proportional to 
one of $\sin (m k_x)$ and $\sin (m k_y)$ with $m = 1,2,\cdots$, 
which vanish at the saddle points 
$(k_x,k_y) = (\pm \pi,0)$ and $(0,\pm \pi)$.

{\it Ruthenate superconductors} --- 
The present model seems to explain 
some of the experimental observations in {\SrRuO}. 
In the experimental result~\cite{Kit10}, 
the transition temperature of the intrinsic state 
is not changed by a perpendicular pressure. 
The theoretical result 
shown in Figs.\,\ref{fig:Tc_tz0Tc15} and \ref{fig:inplane_pw}
indicates that 
this can be explained if the intrinsic state is 
one of 
the \mbox{p-,} sz-p-, cz-p-, and ${\rm d}_{xy}$-wave states. 
The sz- and sz-d-wave states are the second-most likely candidates 
because their $T_{\rm c}$ values weakly depend on $t_z$ 
in the theoretical result. 
\mbox{Table}~\ref{table:compare} lists the order-parameter structures 
and properties of some of the most likely candidates 
for the intrinsic states. 
Among them, only the sz-p and cz-p states exhibit 
horizontal line nodes, 
which are suggested by the field-angle-dependent 
specific-heat measurement~\cite{Kit18}. 
In particular, 
in the spin-triplet state 
$[{\rm cz}$-$({\rm p}_x + i {\rm p}_y)] {\hat \vd}$ 
and 
the spin-singlet state 
${\rm sz}$-$({\rm p}_x + i {\rm p}_y)$, 
the time-reversal symmetry (TRS) is broken, 
which is suggested by muon spin relaxation ($\mu$SR)~\cite{Luk98}, 
where $\vd$ denotes the d-vector and ${\hat \vd} \equiv \vd/|\vd|$. 
Over twenty years, 
it had been considered 
that the absence of the Knight shift~\cite{Ish98} 
supports equal-spin states; 
however, recently, a pronounced drop of $^{17}$O NMR Knight shift 
in the superconducting state was reported~\cite{Pus19,Ish20}. 
This implies that the intrinsic state is an antiparallel spin state 
or at least contains a component of antiparallel spin states, 
where the spin quantization axis is taken in the direction 
parallel to the magnetic field.
The temperature dependence of the upper critical field 
seems to support antiparallel spin states~\cite{Mac08,Mae12}.

\begin{table}[hbtp]
\caption{
Examples of candidates for the intrinsic state 
that is unaffected by the perpendicular pressure. 
${\hat \vx}$, 
${\hat \vy}$, and 
${\hat \vz}$ denote 
the unit vectors in the $x$-, $y$-, and $z$-directions 
in the d-vector space, respectively. 
For the other unlisted candidates, such as the states with 
${\rm p}_x {\hat \vy} \pm {\rm p}_y {\hat \vx}$, 
${\rm p}_x {\hat \vz}$, 
and $({\rm p}_x \pm {\rm p}_y) {\hat \vd}$, 
the properties of the nodes and TRS can easily be found 
from tables in previous studies, 
for example, in Ref.~\citen{Mac03}. 
The only difference is 
the possibility of the factors $\cos k_z$ and $\sin k_z$, 
which add horizontal line nodes to the order parameter. 
}
\label{table:compare}
{\footnotesize 
\begin{center}
\begin{tabular}{cccc}
\hline 
\hline 
Structure of the order parameter       & Spin      & Line nodes & TRS \\ 
\hline 
${\rm p}_x {\hat \vx} \pm {\rm p}_y {\hat \vy}$ 
                                  & triplet & none      & unbroken \\ 
cz-$({\rm p}_x {\hat \vx} \pm {\rm p}_y {\hat \vy})$ 
                                  & triplet & horizontal & unbroken \\ 
${\rm p}_x {\hat \vy} \pm i {\rm p}_y {\hat \vx}$
                                  & triplet & none          & broken \\ 
cz-$({\rm p}_x {\hat \vx} \pm i {\rm p}_y {\hat \vy})$ 
                                  & triplet & horizontal         & broken \\ 
$({\rm p}_x + i {\rm p}_y) {\hat \vd}$   
                                  & triplet & none       & broken   \\ 
$[{\rm cz}$-$({\rm p}_x + i {\rm p}_y)] {\hat \vd}$    
                                  & triplet & horizontal & broken   \\ 
${\rm sz}$-$({\rm p}_x + i {\rm p}_y)$    
                                  & singlet & horizontal & broken   \\ 
${\rm d}_{xy} $                   & singlet & vertical   & unbroken \\ 
\hline 
\hline 
\end{tabular}
\end{center}
}
\end{table}

The observed 3-K phase in {\SrRuO} cannot be among 
\mbox{p-,} sz-p-, cz-p-, and ${\rm d}_{xy}$-wave pairing 
in the present Van Hove scenario, 
because their transition temperatures are almost unaffected by $t_z \ne 0$. 
If any one of them is the 3-K phase, 
$T_{\rm c}$ must be approximately 3\,K for any smaller $t_z$, 
which is inconsistent with the experimental fact. 
Moreover, for the s-, sz-, and sz-d-wave states, 
the enhancement of $T_{\rm c}$ is too weak to be the 3-K phase. 
In contrast, the transition temperatures of 
the \mbox{cz-d-,} \mbox{d-,} and cz-wave states 
are significantly enhanced by $t_z \ne 0$, 
as shown in Fig.\,\ref{fig:Tc_tz0Tc15}, 
and hence, these states are most likely the 3-K phase. 
All of these states are spin-singlet states.

\begin{figure}[htbp]
\begin{center}
\begin{tabular}{c}
\includegraphics[width=5.5cm]{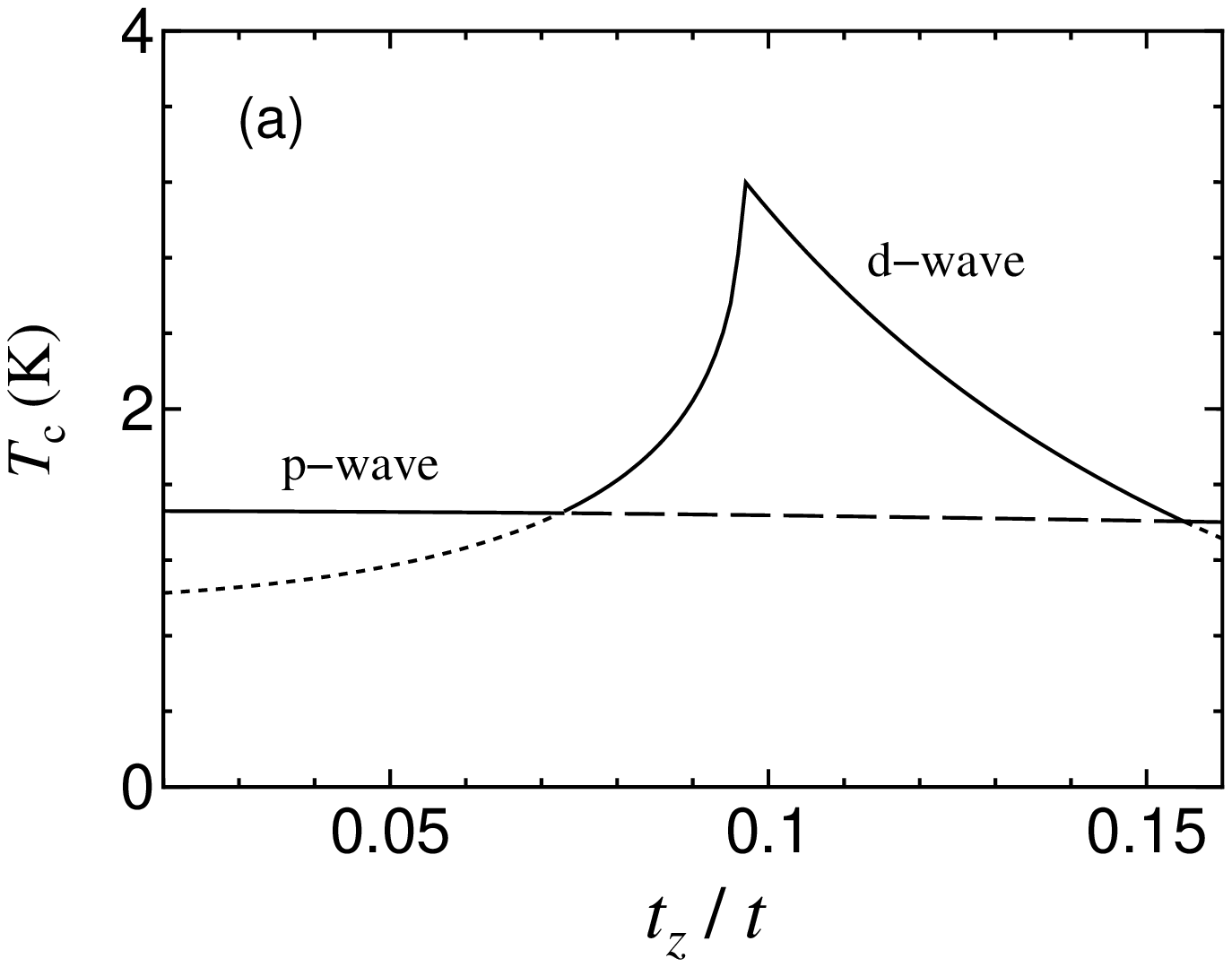}
\\[2pt]
\includegraphics[width=5.5cm]{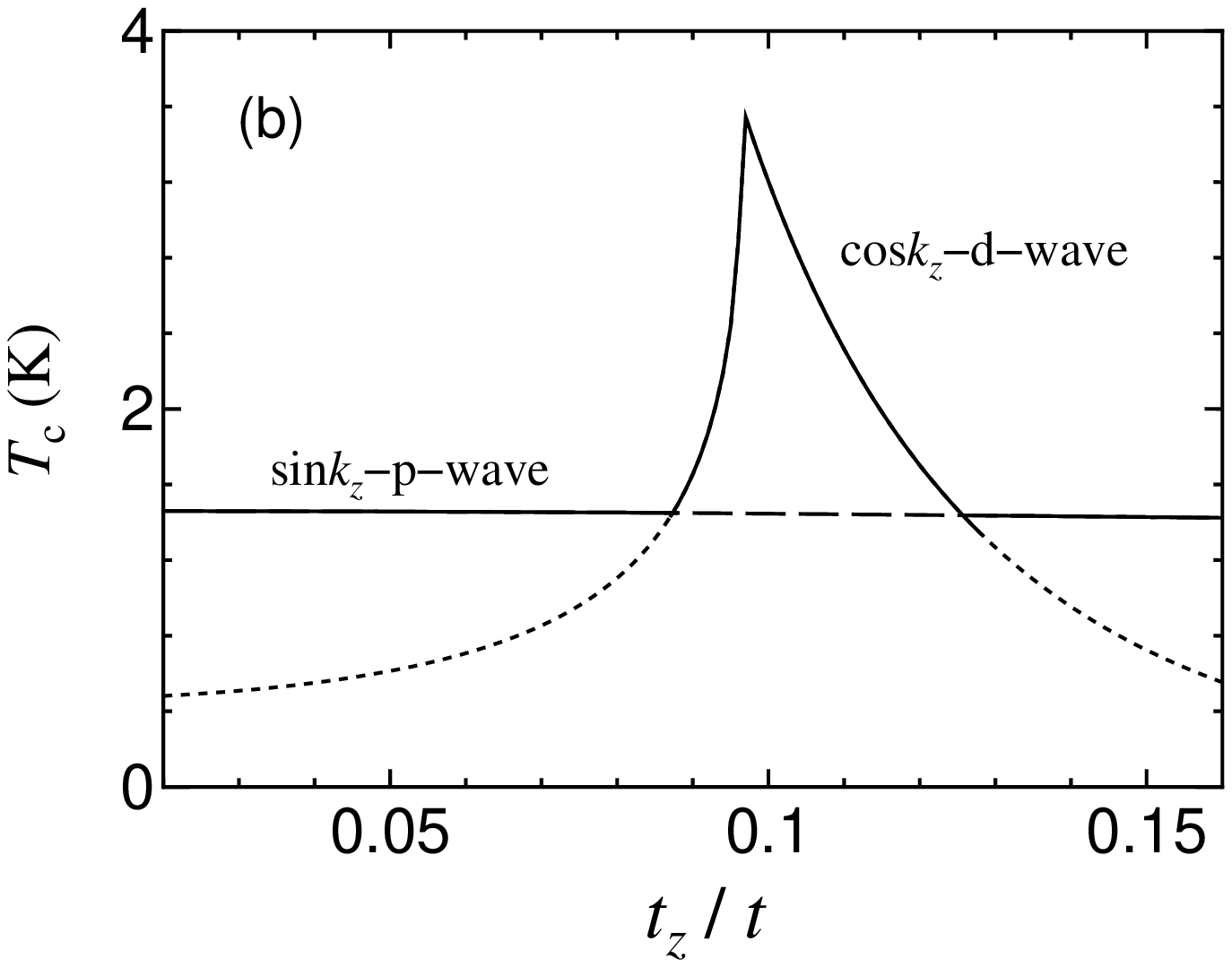}
\\[-12pt]
\end{tabular}
\end{center}
\caption{
Transition temperatures when 
$n = 0.9$ and $\omega_{\rm c} = 300~{\rm K}$. 
(a) When p-wave and d-wave pairing interactions coexist. 
$g_{\rm p} =  0.91 t$ and $g_{\rm d} = 0.34 t$ are assumed. 
(b) When sz-p-wave and cz-d-wave pairing interactions coexist. 
$g_{\rm sz-p} =  0.91 t$ and $g_{\rm cz-d} = 0.30 t$ are assumed. 
}
\label{fig:2ndtrns}
\end{figure}

Figure~\ref{fig:2ndtrns} depicts the Van Hove scenario for {\SrRuO} 
under perpendicular pressures. 
Since the transition temperature given in \eq.{eq:Tcexpress} 
is the instability temperature of 
the normal state, 
only the highest one is realized at each $t_z$. 
Therefore, the candidates for the intrinsic phase 
are the p-, \mbox{cz-p-}, and sz-p-wave states, 
and 
the candidates for the higher-temperature phase 
are the d-, cz-, and cz-d-wave states. 
Figures~\ref{fig:2ndtrns} (a) and (b) 
present the results of two examples of combinations 
of the intrinsic and higher-temperature phases, 
i.e., 
the p- and d-wave states 
and the sz-p- and cz-d-wave states, respectively. 
In both cases, 
because the higher-temperature phases 
(d- and cz-d-wave states) 
have more nodes than the intrinsic phases 
(chiral p- and chiral sz-p-wave states, respectively), 
the transitions presented by the dotted curves 
must be completely suppressed, 
whereas the dashed curves might survive as 
approximate transition temperatures to mixed states 
or approximate first-order transition temperatures 
to the low-temperature phases.


{\it Conclusion} --- 
In conclusion, 
it was shown that the superconducting transition temperature 
can be strongly enhanced by uniaxial pressures perpendicular 
to the most conductive layers in quasi-two-dimensional superconductors 
off half-filling 
because of a vestigial Van Hove singularity. 
We examined this effect for various types of pairing states 
including those induced by interlayer pairing. 
Among them, the enhancement is the largest 
for the interlayer d-wave state 
with 
$\Delta_{\vk} \propto \cos k_z (\cos k_x - \cos k_y)$, 
and it is also large for the d-wave state 
with 
$\Delta_{\vk} \propto \cos k_x - \cos k_y$. 
In contrast, 
this effect does not exist 
for the interlayer and intralayer p-wave states, 
because $\sin (m k_x)$ and $\sin (m k_y)$ vanish 
at $(k_x,k_y) = (\pm \pi,0)$ and $(0,\pm \pi)$. 
These behaviors are consistent with experimental observations in 
{\SrRuO} under perpendicular pressures~\cite{Kit10}, 
if we assume that the higher-temperature phase is 
one of the intralayer and interlayer spin-singlet 
d- and s-wave states 
and the intrinsic 1.5-K phase is one of 
the intralayer and interlayer p-wave states. 
The interlayer p-wave states can be either spin-singlet 
or spin-triplet states depending on the factors $\cos k_z$ and $\sin k_z$, 
respectively.

As future studies, 
the structures of the mixed states below the second 
(lower) transition temperature when the 3-K phase occurs 
and 
the superconductivity under uniaxial pressures in the other directions 
will be examined in separate papers. 
For a close comparison with the observed facts in {\SrRuO}, 
details of the Fermi-surface structures 
of all $\alpha$, $\beta$, and $\gamma$ bands 
may need to be incorporated.

\mbox{}






\begin{thebibliography}{99}
\bibitem{Noterhopara} 
  Expressions for $\rho_{\parallel}$ are presented 
  in previous papers, such as \mbox{Refs.~\citen{Shi87} and \citen{Shi88}.}
\bibitem{Notespps}  
  An example in which a linear combination of basis functions 
  is localized near the Fermi surface for the most stable state 
  is presented in Ref.~\citen{Shi89}.
\bibitem{Notegalpha}
  We assume the coupling constant of the principal basis $\gamma_{\alpha}$ 
  so that $T_{\rm c} \approx 1.5~{\rm K}$ at $t_z = 0$: 
  $g_{\alpha} = 0.81 t$ if $\alpha$ is s, sz, or cz; 
  $g_{\alpha} = 0.365 t$ if $\alpha$ is d, sz-d, or cz-d; 
  $g_{\alpha} = 0.91 t$ if $\alpha$ is p; 
  and 
  $g_{\alpha} = 0.69 t$ if $\alpha$ is ${\rm d}_{xy}$.
\end{thebibliography}
\end{document}